\pdfoutput=1

\documentclass[12pt]{article}
\usepackage{graphicx}
\usepackage{multicol}
\usepackage{subfig}
\usepackage{lineno}

\newcommand\pubnumber{ATL-PHYS-PROC-2022-117}
\newcommand\pubdate{\today}

\def\institute{Comenius University, Bratislava, Slovakia}

\def\Title#1{\begin{center} {\Large #1 } \end{center}}
\def\Author#1{\begin{center}{ \sc #1} \end{center}}
\def\Address#1{\begin{center}{ \it #1} \end{center}}

\newcommand\pubblock{\rightline{\begin{tabular}{l} \pubnumber\\
         \pubdate  \end{tabular}}}
\newenvironment{Abstract}{\begin{quotation}  }{\end{quotation}}
\newenvironment{Presented}{\begin{quotation} \begin{center} 
             PRESENTED AT\end{center}\bigskip 
      \begin{center}\begin{large}}{\end{large}\end{center} \end{quotation}}

\def\beq{\begin{equation}}
\def\eeq#1{\label{#1}\end{equation}}
\def\eeqn{\end{equation}}

\def\beqa{\begin{eqnarray}}
\def\eeqa#1{\label{#1}\end{eqnarray}}
\def\eeqan{\end{eqnarray}}

\let\bar=\overbar

\def\Dslash{\not{\hbox{\kern-4pt $D$}}}
\def\dslash{\not{\hbox{\kern-2pt $\del$}}}

\def\msb{{\bar{\ssstyle M \kern -1pt S}}}

\def\ttbar{$t\bar{t}$}
\def\mtt{$m_{t\bar{t}}$}

\begin{document}
\begin{titlepage}
\pubblock

\vfill
\Title{Charge asymmetry in top-pair production\\\vspace{0.25cm} with the ATLAS detector}
\vfill
\Author{ Barbora Eckerova}
\vspace{-1cm}
\Address{\institute}
\begin{center}\sc on behalf of the ATLAS Collaboration \end{center}
\vfill
\begin{Abstract}
Measurement of the inclusive and differential top-quark charge
asymmetry is carried out using full Run 2 data from proton-proton collisions at a
center-of-mass energy of $13$~TeV collected by the ATLAS detector.
The single-lepton and dilepton \ttbar~decay channels are combined.
Distorting detector effects are removed using fully Bayesian
unfolding.
In the dilepton decay channel, the leptonic charge asymmetry is
determined. Results from the measurements are compared with the
Standard Model prediction. The excess of charge asymmetry from zero charge-asymmetry hypothesis at the level of $4.7$ standard deviations is observed.
\end{Abstract}
\vfill
\begin{Presented}
$15^\mathrm{th}$ International Workshop on Top Quark Physics\\
Durham, UK, 4--9 September, 2022
\end{Presented}
\vfill
\let\thefootnote\relax\footnotetext{Copyright 2022 CERN for the benefit of the ATLAS Collaboration.	CC-BY-4.0 license.}
\end{titlepage}
\def\thefootnote{\fnsymbol{footnote}}
\setcounter{footnote}{0}

\section{Introduction}

The top-quark charge asymmetry is one of the features of top-quark pair production. The cross section formula of \ttbar~production is not symmetric under the exchange of final-state top quark and top anti-quark. The asymmetric contribution arises after the inclusion of next-to-leading order diagrams of \ttbar~production. 

\begin{figure}[htb]
	\centering
		\includegraphics[width=0.6\textwidth,trim={0cm 25cm 0cm 0cm }]{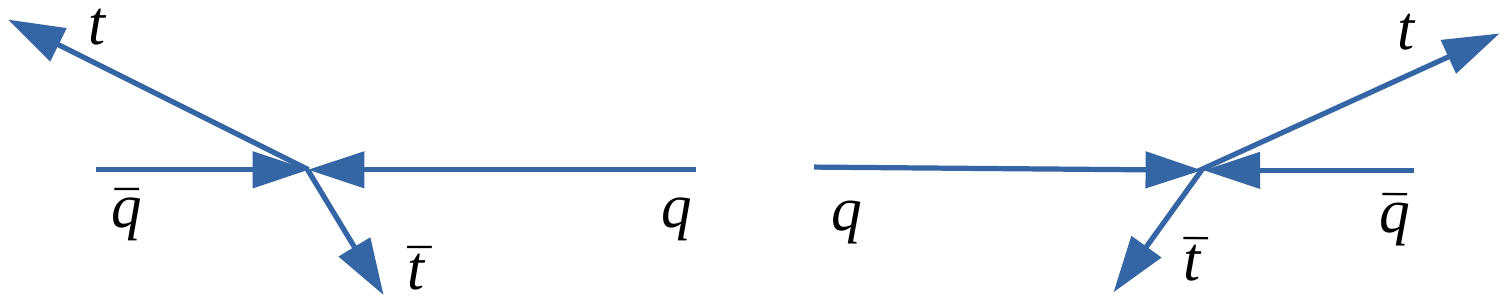}
		\includegraphics[width=0.3\textwidth]{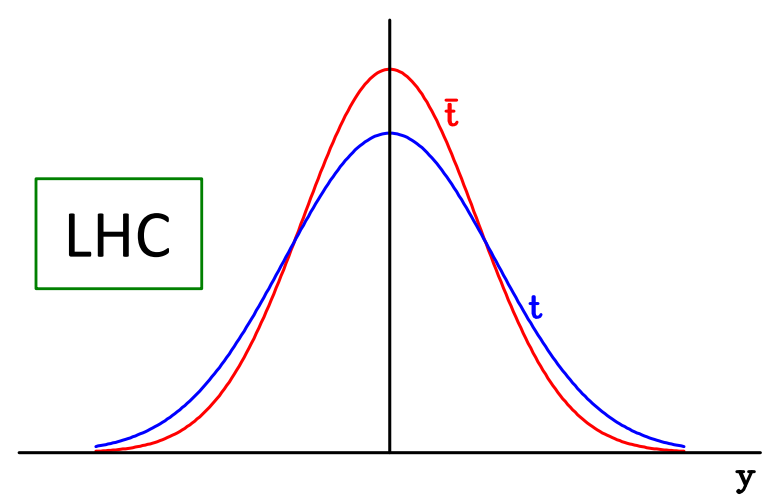}
\caption{(Left) Preferred direction of the top quark and top anti-quark in $q\bar q$ annihilation. Top quarks are generally more longitudinally boosted than top anti-quarks. (Right) An illustration of rapidity distributions of top quark and top anti-quark as measured by the LHC~\cite{LHC}.}
	\label{fig:rapidity_LHC}
\end{figure}

The charge asymmetry is not present in top-quark pair production initiated by gluon fusion. The largest contribution originates from interactions of initial-state quarks ($q$) and anti-quarks ($\bar q$)~\cite{Rodrigo1,Rodrigo2}. Due to the charge asymmetry, top quarks are preferentially produced in the direction of initial-state $q$, and top anti-quarks in the direction of initial-state $\bar q$ as is illustrated by Fig.~\ref{fig:rapidity_LHC} (left).
Hence, due to a longitudinal momentum imbalance of initial-state $q$ and $\bar q$, the charge asymmetry is manifested by different rapidity distributions of top quark ($t$) and top anti-quark ($\bar t$), as shown in Fig.~\ref{fig:rapidity_LHC} (right).

The charge asymmetry is based on computing absolute rapidity difference of $t$ and $\bar t$, $\Delta |y|=|y_t| - |y_{\bar t}|$. The sign of absolute rapidity difference signifies the direction of $t$.  
The charge asymmetry is evaluated as
\begin{center}
$A_C^{t\bar t} = \frac{N(\Delta |y| >0)- N(\Delta |y|< 0)}{N(\Delta |y| >0)+ N(\Delta |y| < 0)}$.
\end{center}
Specifically for dilepton \ttbar~decay channel, the leptonic charge asymmetry $A_C^{\ell \bar \ell}$ is defined similarly. Instead of $\Delta |y|$, absolute pseudorapidity difference $\Delta |\eta_{\ell \bar \ell}|=|\eta_\ell|-|\eta_{\bar \ell}|$ of two leptons originating from top-quark pair decay is used.

In proton-proton collisions, \ttbar~pairs are produced in approximately $90$\% of cases by charge symmetric gluon-gluon fusion. Therefore, the size of the charge asymmetry is below $1$\%, so the measurement of the charge asymmetry is challenging.

The enhancement of the charge asymmetry is expected in specific parts of the phase space. 
Hence, the charge asymmetry is studied as a function of three different observables: \mtt, longitudinal boost of \ttbar~pair along the beam axis $\beta_{z,t\bar t}$ and transverse momentum of \ttbar~pair $p_{T,t\bar t}$.

\section{Event selection}

The inclusive and differential charge asymmetry is measured using data collected by the ATLAS detector~\cite{ATLAS} at $\sqrt{s}=13$~TeV, which correspond to $139~\textrm{fb}^{-1}$. The single-lepton and dilepton \ttbar~decay channels are both used for measurement.

For single-lepton events, two topology types are exploited - resolved and boosted. In resolved topology, reconstruction of \ttbar~events is provided by using Boosted Decision Tree (BDT).  
Data in dilepton decay channel are divided according to flavor of leptons in same-flavor ($ee+\mu\mu$) and opposite flavor ($e\mu$) regions. The \ttbar~system is reconstructed using Neutrino Weighting~\cite{NW}.
 Events in both decay channels are further divided according to $b$-jet multiplicity to $1~b$-tag exclusive and $2~b$-tag inclusive events.

\section{Fully Bayesian unfolding}

The measured distribution of $\Delta|y|$ or $\Delta|\eta|$ is unfolded to the parton level using fully Bayesian unfolding (FBU)~\cite{FBU}. Distorting effects of the detector, like limited acceptance and resolution, are corrected by the unfolding.

The FBU exploits Bayesian inference, $p(T|D)\propto\mathcal{L}(D|T) \cdot \pi(T)$, to obtain posterior probability distribution $p(T|D)$ of true distribution $T$ using measured data $D$ and simulated response matrix $\mathcal{M}$.
 Systematic uncertainties are embedded in the likelihood function $\mathcal{L}$ by implementation of nuisance parameters $\theta$. After marginalization of nuisance parameters (NPs), obtained constraints together with correlations of NPs lead to the reduction of the unfolded uncertainty.

Uniform prior probability $\pi(T)$ for bins of true spectrum is chosen, whereas Gaussian probability terms are used for systematic NPs.

\section{Results}

\begin{figure}[htb]
	\centering
			\includegraphics[width= 0.43\textwidth]{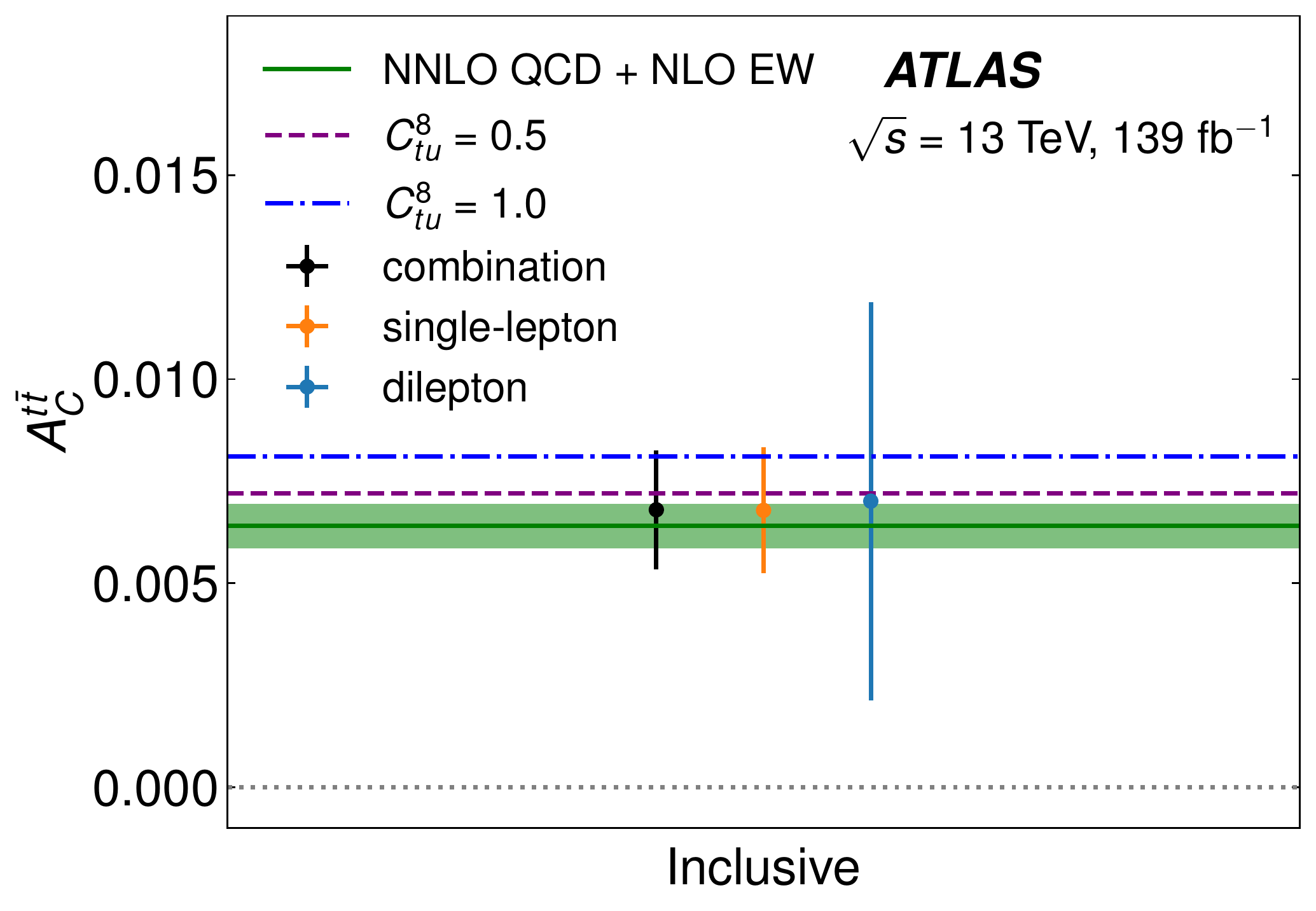}
			\includegraphics[width=0.43\textwidth]{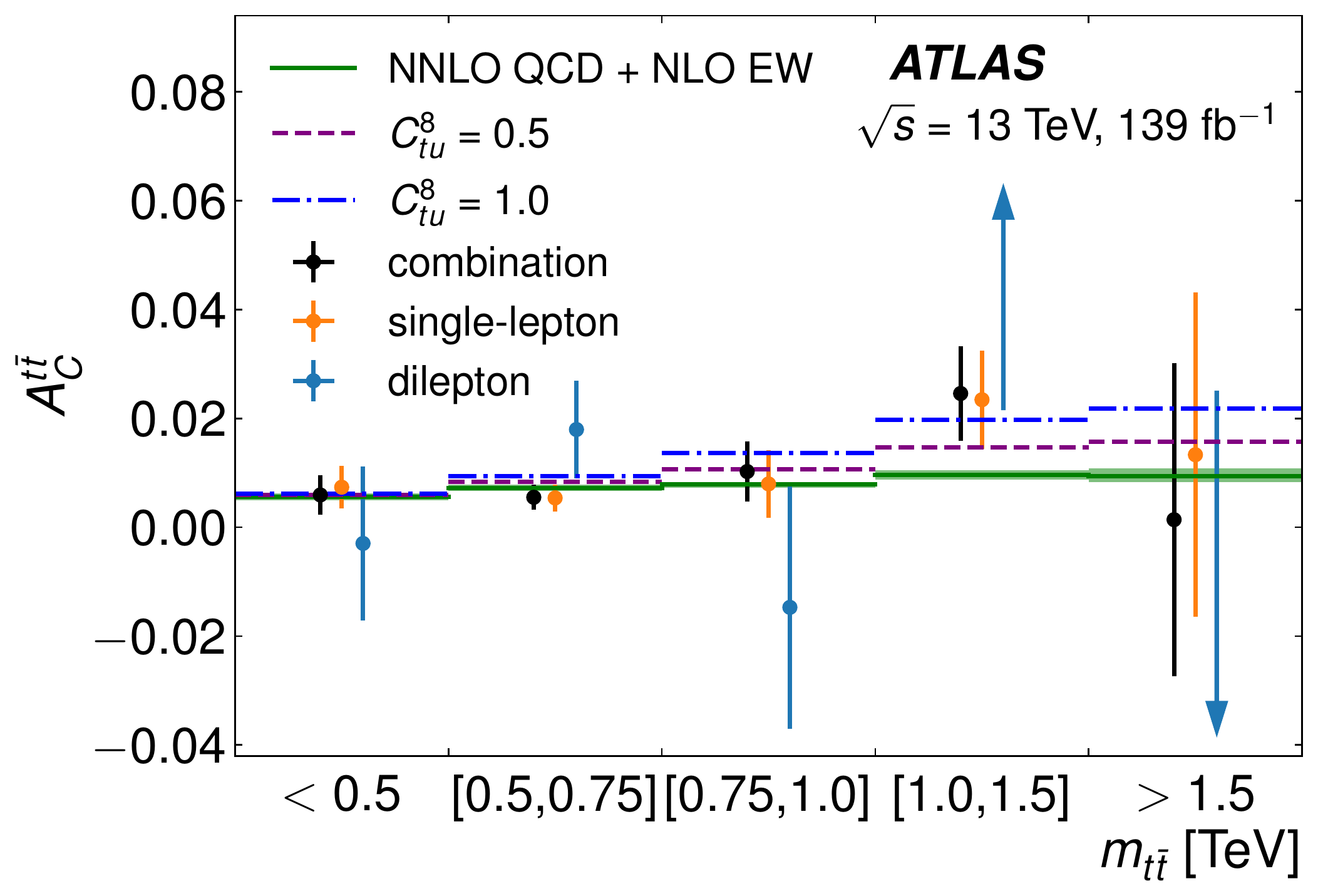}
	\caption{Unfolded $A_C^{t\bar t}$ in inclusive (left) and \mtt~differential measurements (right)~\cite{TCA13_comb}.}%
	\label{fig:ac}
\end{figure}

The charge asymmetry value is unfolded inclusively and differentially as a function of \mtt, $\beta_{z,t\bar t}$ and $p_{T,t\bar t}$. Single-lepton and dilepton events are unfolded separately, but also in combination. Similarly, leptonic charge asymmetry measurement is performed for events in the dilepton channel inclusively and in a differential measurement as a function of $m_{\ell \bar \ell}$, $\beta_{z,\ell\bar \ell}$ and $p_{T,\ell\bar \ell}$. In Fig.~\ref{fig:ac}, unfolded charge asymmetry values for inclusive and \mtt~differential measurements are summarized for combination of single-lepton and dilepton data, but also for events for each decay channel separately. Inclusive $A_C^{t\bar t}$ for the combination measurement is significantly different from the hypothesis of zero charge asymmetry. The non-zero excess of $A_C^{t\bar t}$ is $4.7$ standard deviations. Therefore, the result provides evidence of charge asymmetry in \ttbar~production at the LHC.

The leptonic charge asymmetry is measured to be $0.0054\pm0.0026$ in the inclusive measurement.
Results from all measurements, either $A_C^{t\bar t}$ or $A_C^{\ell\bar \ell}$, are in good agreement with the SM prediction calculated at NLO in EW theory and either to NNLO or NLO in QCD, for \ttbar~and leptonic charge asymmetries, respectively~\cite{NNLO_ttbar,NNLO_leptonic}.

\section{EFT interpretation}

The charge asymmetry measurement has a potential to provide extra information into global SMEFT fits.

\begin{figure}[hbt]
	\centering
	\begin{minipage}[c]{0.4\textwidth}
		\centering
			\includegraphics[height=6.8cm]{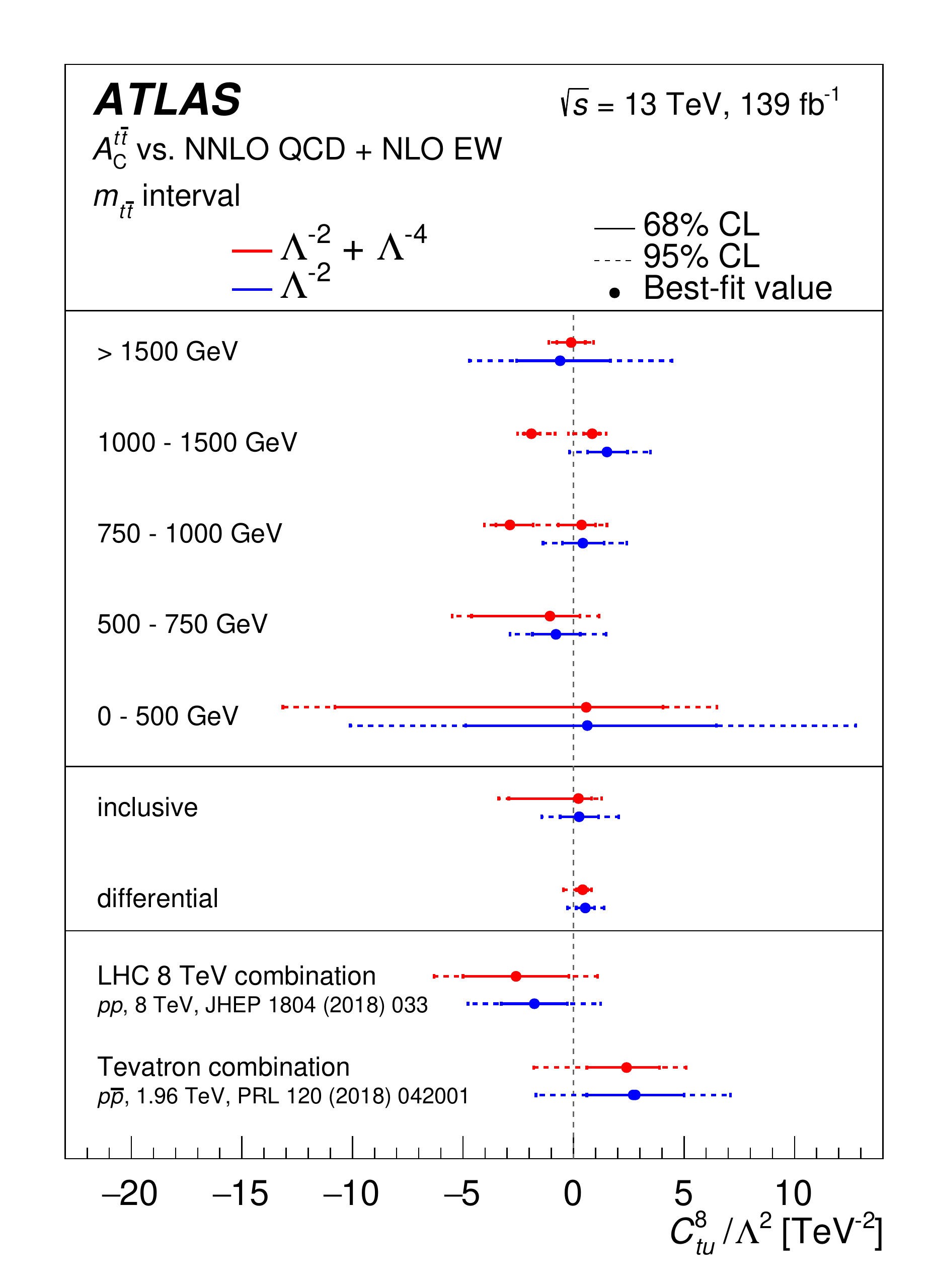}
	\end{minipage}\hspace{1cm}
	\begin{minipage}[c]{0.4\textwidth}
		\centering
		\includegraphics[height=5.8cm]{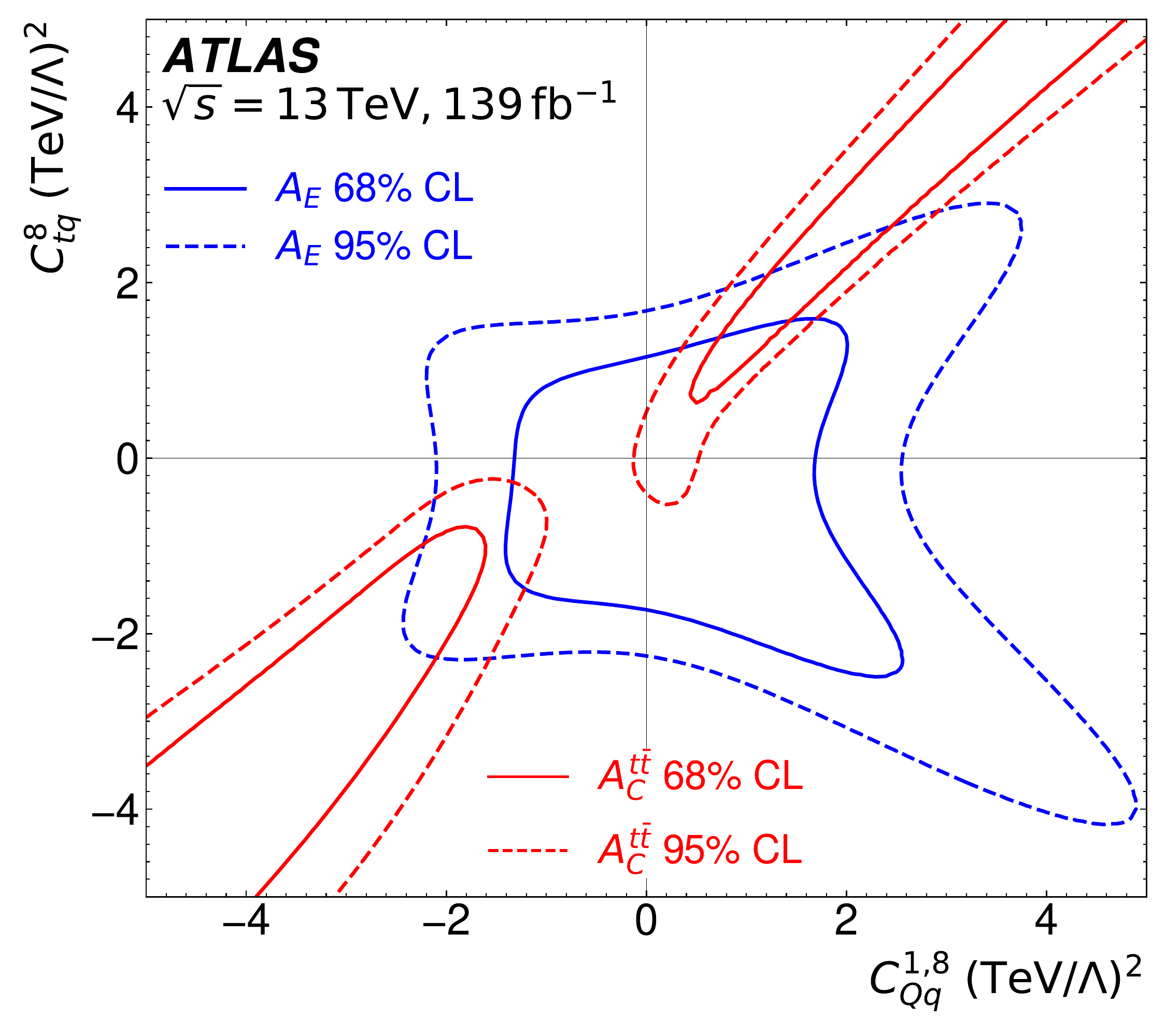}
	\end{minipage}
	\caption{(Left) Bounds on the Wilson coefficient $C_{tu}^{8}$ derived from inclusive and \mtt~differential $A_C^{t\bar t}$ measurement. (Right)  Bounds from \mtt~differential $A_C^{t\bar t}$ measurement and limits derived from $A_E$ measurement~\cite{AE,TCA13_comb}.}
	\label{fig:AE_Ac_EFT}
\end{figure}

Inclusive and \mtt~differential measurements of $A_C^{t\bar t}$ are used to derive constraints on various Wilson coefficients. In Fig.~\ref{fig:AE_Ac_EFT} (left), bounds obtained from individual \mtt~differential bins together with constraint derived from the inclusive measurement are summarized. The most stringent limit of the $C^{8}_{tu}$ coefficient is obtained using information from all \mtt~differential bins. The combined constraint is more than a factor of two more stringent than the bound from the inclusive measurement. The sensitivity increases with higher values of \mtt.~Constraints derived from previous measurements are presented in Fig.~\ref{fig:AE_Ac_EFT} (left) for reference.

The measurement of energy asymmetry $A_E$~\cite{AE} is complementary to charge asymmetry measurement in context of EFT interpretation. This is illustrated in Fig.~\ref{fig:AE_Ac_EFT} (right), where limits derived from both measurements ($A_E$ and $A_C^{t\bar t}$) for coefficients $C^{1,8}_{Qq}$ and $C^8_{tq}$ are plotted. The charge asymmetry measurement does not supply any information in one part of parameter space (left upper panel of the plot). The energy asymmetry measurement is sensitive to those values of Wilson coefficients. Therefore, common plots of bounds for EFT operators give more information.

\end{document}